\documentclass[runningheads]{llncs}
\usepackage[T1]{fontenc}
\usepackage{graphicx} 
\usepackage{amsmath}
\usepackage{cite}
\usepackage{amssymb}
\usepackage{booktabs}
\usepackage{adjustbox}
\usepackage{colortbl}
\usepackage[table]{xcolor}
\usepackage{multirow}
\usepackage{array}
\usepackage{float}
\usepackage{hyperref}
\usepackage{orcidlink} 

\begin{document}
\title{Global and Local Contrastive Learning for Joint Representations from Cardiac MRI and ECG}
\titlerunning{Patient and Temporal Alignment Contrastive Learning}
% If the paper title is too long for the running head, you can set
% an abbreviated paper title here
%
 \author{Alexander Selivanov\inst{1}\orcidlink{0009-0005-8945-4547} \and
 Philip M\"uller\inst{1,2}\orcidlink{0000-0001-8186-6479} \and
 \"Ozg\"un Turgut\inst{1,2}\orcidlink{0009-0002-8704-0277} \and
 Nil Stolt-Ans\'o\inst{1,4}\orcidlink{0009-0001-4457-0967} \and 
Daniel R\"uckert\inst{1,2,3,4}\orcidlink{0000-0002-5683-5889}}
\authorrunning{A. Selivanov et al.}
% First names are abbreviated in the running head.
% If there are more than two authors, 'et al.' is used.
%
% index{Selivanov, Alexander}
% index{M\"uller, Philip}
% index{Turgut, \"Ozg\"un}
% index{Stolt-Ans\'o, Nil}
% index{R\"uckert, Daniel}

\institute{Chair for AI in Healthcare and Medicine, Technical University of Munich (TUM) and TUM University Hospital, Munich, Germany \and
School of Medicine, Klinikum rechts der Isar, TUM, Germany \and
Department of Computing, Imperial College London, UK  \and
Munich Center for Machine Learning (MCML), Munich, Germany \\
    \email{alexander.selivanov@tum.de}}

% \author{Anonymized Authors}  %% Added for anonymized MICCAI 2025 submission
% \authorrunning{Anonymized Author et al.}
% \institute{Anonymized Affiliations \\
%     \email{email@anonymized.com}}

%
\maketitle              % typeset the header of the contribution
\begin{abstract}
An electrocardiogram (ECG) is a widely used, cost-effective tool for detecting electrical abnormalities in the heart. However, it cannot directly measure functional parameters, such as ventricular volumes and ejection fraction, which are crucial for assessing cardiac function. Cardiac magnetic resonance (CMR) is the gold standard for these measurements, providing detailed structural and functional insights, but is expensive and less accessible. To bridge this gap, we propose PTACL (\textbf{P}atient and \textbf{T}emporal \textbf{A}lignment \textbf{C}ontrastive \textbf{L}earning), a multimodal contrastive learning framework that enhances ECG representations by integrating spatio-temporal information from CMR. PTACL uses global patient-level contrastive loss and local temporal-level contrastive loss. The global loss aligns patient-level representations by pulling ECG and CMR embeddings from the same patient closer together, while pushing apart embeddings from different patients. Local loss enforces fine-grained temporal alignment within each patient by contrasting encoded ECG segments with corresponding encoded CMR frames. This approach enriches ECG representations with diagnostic information beyond electrical activity and transfers more insights between modalities than global alignment alone, all without introducing new learnable weights. We evaluate PTACL on paired ECG-CMR data from 27,951 subjects in the UK Biobank. Compared to baseline approaches, PTACL achieves better performance in two clinically relevant tasks: (1) retrieving patients with similar cardiac phenotypes and (2) predicting CMR-derived cardiac function parameters, such as ventricular volumes and ejection fraction. Our results highlight the potential of PTACL to enhance non-invasive cardiac diagnostics using ECG. The code is available at: \url{https://github.com/alsalivan/ecgcmr}

\end{abstract}
\section{Introduction}
Cardiovascular diseases (CVD) remain a leading cause of mortality worldwide \cite{who-mortality-database}. Electrocardiograms (ECGs) are a non-invasive and cost-effective tool for detecting electrical heart abnormalities (e.g., arrhythmias \cite{Ansari2023}). However, ECGs can't directly assess structural and functional cardiac properties like ventricular volumes, myocardial mass, or ejection fraction. In contrast, cardiac magnetic resonance (CMR) is the gold standard for these assessments \cite{Leiner2020}, but it is expensive, time-consuming, and requires specialized expertise, limiting its accessibility.

To integrate information from both modalities, researchers have applied contrastive learning to align ECG and CMR embeddings globally \cite{baseline1, mmcl, CMAE}, pulling same-patient representations closer while pushing apart different patients. However, these approaches treat ECG and CMR as holistic representations, overlooking finer temporal relationships. Since CMR captures dynamic sequences across the cardiac cycle while ECG records continuous electrical activity, global alignment alone lacks the temporal precision needed to effectively integrate both modalities.

\begin{figure}[ht!]
    \centering
    \includegraphics[width=0.99\textwidth]{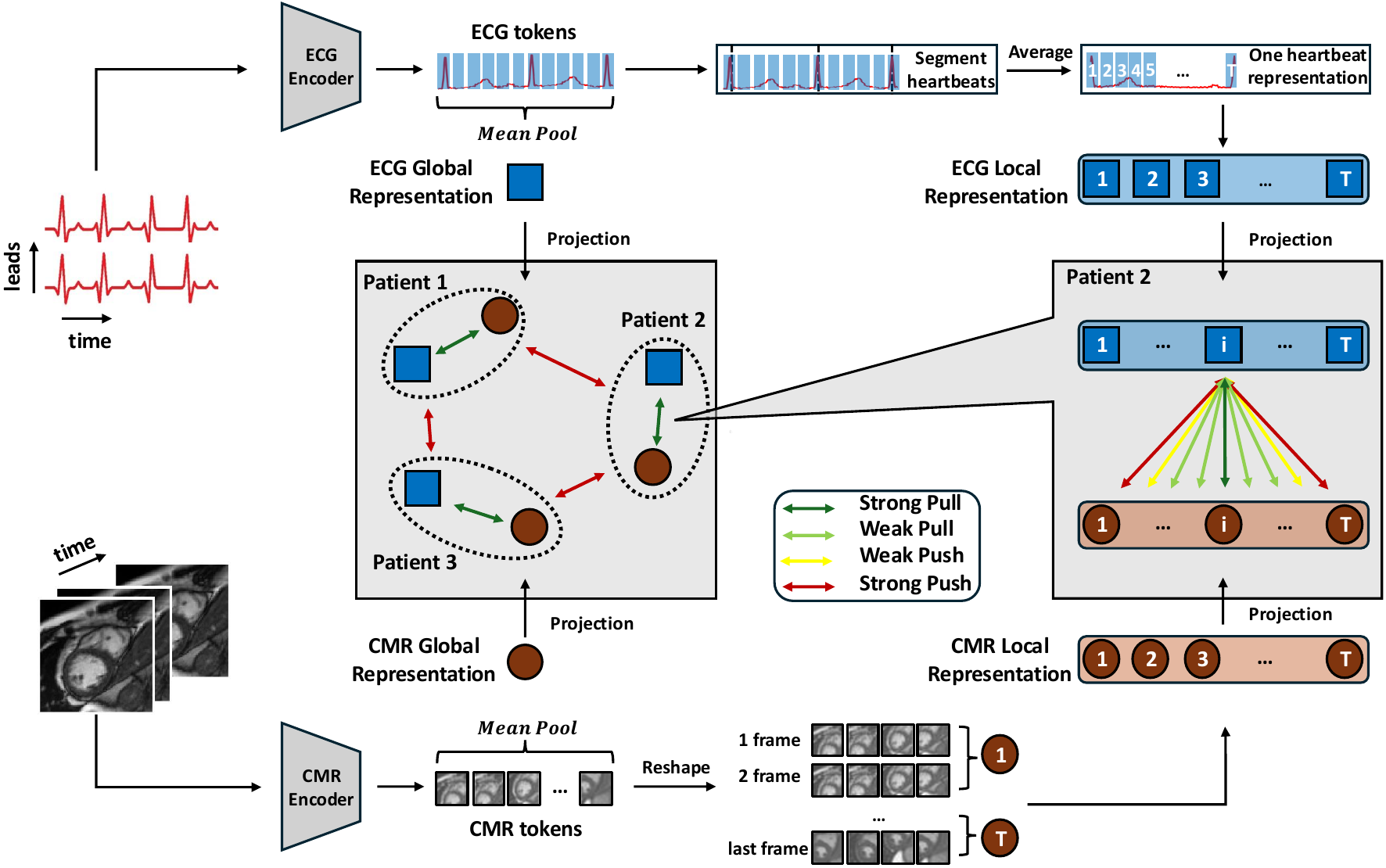}
    \caption{Overview of PTACL. We train a multimodal model on pairs of ECG and CMR using a combination of a global contrastive loss (left square) and a local contrastive loss (right square). The global loss aligns patient-level ECG and CMR embeddings by pulling representations from the same patient closer together while pushing apart those from different patients. The local loss enforces fine-grained temporal alignment by contrasting ECG-derived segments—encoded from a single-heartbeat representation—with their corresponding CMR frames. In the figure, representations from 1 to T start at the end-diastole (ED) phase. Overlays in the tokens (CMR and ECG) are for visualization purposes only.}
    \label{fig:method}
\end{figure}

To address this limitation,  we introduce PTACL (\textbf{P}atient and \textbf{T}emporal \textbf{A}lign\-ment \textbf{C}ontrastive \textbf{L}earning), a contrastive learning framework that combines global and local alignment. The global loss aligns ECG and CMR embeddings at the patient level, while the local loss enforces fine-grained temporal alignment by contrasting ECG-derived segments—encoded tokens from a single heartbeat representation—with their corresponding CMR frame representations at the same cardiac phase (Fig.~\ref{fig:method}). Notably, the local alignment is completely parameter-free, introducing no additional learnable parameters. Our method is fully self-supervised and is trained on paired ECG-CMR data from 27,951 subjects in the UK Biobank \cite{ukbiobank}. We evaluate PTACL on patient retrieval and phenotype regression, assessing how well ECG embeddings capture CMR-derived cardiac function. Our results demonstrate that local contrastive alignment improves ECG representations, enhancing their diagnostic utility for assessing cardiac structure and function.

\section{Related Works}
Self-supervised learning has advanced both single modality and multimodal representation learning. For single modality tasks, masked autoencoders (MAE) \cite{image_mae} have proven highly effective in learning meaningful representations from unlabeled data. They have been widely applied to signals such as ECG \cite{MaeFe, mae_ecg, mae_ecg_2, mae_ecg_3, mae_ecg_apple, Yang2024}, speech, and audio \cite{audiomae}, as well as medical imaging, where 3D MAEs \cite{videomae, 3dmae} learn spatial-temporal features from volumetric scans \cite{heart2dt, amaes, swinunetr}.

In multimodal learning, most contrastive approaches are inspired by CLIP \cite{clip}, which aligns per-patient representations from different modalities in a shared embedding space, pulling similar samples together while pushing apart unrelated ones. This framework has been applied in medical domains, such as ECG-CMR integration \cite{baseline1, mmcl, CMAE} and X-ray–text association \cite{convirt, medclip, Chexzero}, enabling zero-shot classification and improved diagnostic insights. However, these methods primarily focus on global alignment, treating each modality as a single representation and potentially overlooking fine-grained local interactions.

To address this, recent work has started exploring local contrastive learning for medical applications. For example, GLORIA \cite{gloria} applies a global-local learning scheme to image-text pairs, while Seibold et al. \cite{Seibold_2022} refine CLIP by contrasting local image features with radiology reports, and LoVT \cite{local_learn} aligns image subregions with text embeddings. These methods mirror broader trends in vision-language research \cite{xclip, taco, filip}.

\section{Methods}
Our framework consists of three phases: single modality pre-training (SMP) (Sec.~\ref{sec:single_modality}), multimodal contrastive learning (Sec.~\ref{sec:multi_modality}), and inference. First, we pre-train separate ECG and CMR encoders using MAE \cite{image_mae} on unlabeled data. Then, we jointly train both encoders with substantial augmentations, applying a global contrastive loss for patient-level alignment and a local contrastive loss for fine-grained temporal alignment. The local loss contrasts ECG-derived time segments—fixed length segments from a single heartbeat representation of multi-lead ECG—with corresponding encoded frames from CMR at the same cardiac phase. Finally, we evaluate the pre-trained ECG encoder on multiple downstream tasks.

\subsection{Pre-training for ECG and CMR} 
\label{sec:single_modality}
To integrate ECG and CMR, each modality first learns meaningful representations independently through single modality pre-training (SMP). This step enables the encoders to capture modality-specific features before multimodal contrastive learning. Following previous works \cite{MaeFe, audiomae, videomae}, we pre-train the ECG and CMR encoders independently on each modality, using the MAE framework \cite{image_mae} in both cases.

We embed the multi-lead ECG signals using a 1D convolution over the time dimension with kernel size \( p_t \), transforming each lead into a sequence of non-overlapping time-based tokens. The convolution is applied on each lead independently, with weights shared across all leads. To learn temporal and lead structure, we add learnable positional embeddings before randomly masking a portion of tokens.

For CMR, we treat the 2D+T short-axis sequence as a multi-frame input and apply a 3D convolution with kernel size \([t, p, p]\), where \( t \) defines the temporal window and \( p \) controls the spatial patch size. This generates a sequence of spatio-temporal tokens, each representing a localized region of the heart across time. To learn spatial and temporal structure, we add learnable positional embeddings before masking.

In both cases, a transformer encoder processes only the visible tokens, while a lightweight transformer decoder reconstructs the masked tokens. The Mean Squared Error (MSE) loss is computed only on the masked tokens, comparing the reconstructed output to the original uncorrupted input.

\subsection{Multimodal Contrastive Pre-Training}
\label{sec:multi_modality}
To learn robust cross-modal representations, we employ two contrastive losses: a global loss for patient-level alignment and a local loss for fine-grained temporal alignment. The global loss ensures that embeddings from the same patient are similar while pushing apart embeddings from different patients. The local loss further refines alignment by matching ECG-derived temporal embeddings with corresponding CMR frames. ECG local embeddings consist of \( T \) segments, each spanning \( p_t \) time steps from a single heartbeat representation of multi-lead ECG, while CMR embeddings represent encoded frames from the cardiac sequence.

\subsubsection{Global Contrastive Loss:}
For a batch \( B \), let \( z_{\mathcal{E}}, z_{\mathcal{C}} \) be the ECG and CMR embeddings, obtained via average pooling over token representations followed by separate projection layers. We enforce cross-modal alignment using the InfoNCE loss, following \cite{clip}. The loss is defined symmetrically for $\text{ECG} \!\to\! \text{CMR}$ and $\text{CMR} \!\to\! \text{ECG}$ directions. The $\text{ECG} \!\to\! \text{CMR}$ contrastive loss is:

\begin{equation}
\mathcal{L}_{\text{ECG} \to \text{CMR}} = \frac{1}{B} \sum_{i=1}^B -\log \frac{\exp \left( \cos \left( z_{\mathcal{E}}^{(i)}, z_{\mathcal{C}}^{(i)} \right) / \tau \right) }{\sum_{j=1}^B \exp \left( \cos \left( z_{\mathcal{E}}^{(i)}, z_{\mathcal{C}}^{(j)} \right) / \tau \right) },
\end{equation}

where \( \tau \) is the temperature parameter, \( z_{\mathcal{E}} \) and \( z_{\mathcal{C}} \) are the projected embeddings, and \( \cos(\cdot, \cdot) \) denotes cosine similarity. 
The final global contrastive loss \( \mathcal{L}_{\text{global}} \) is a weighted combination of both directions.

\subsubsection{Local Contrastive Loss:}
CMR acquisition is guided by ECG gating, where imaging data is continuously recorded over multiple heartbeats while simultaneously capturing the ECG signal. The detected R peaks are later used to align and reconstruct a single representative heartbeat cycle. This physiological alignment motivates our local contrastive loss formulation.

\paragraph{Local Embedding Formation:}
\label{sec:local_embeddings_formation}
For CMR, each frame is divided into spatial tokens representing \( p \times p \) image patches and temporal tokens capturing sequences of \( t \) frames. Local embeddings are obtained by averaging spatial tokens within each temporal token, producing \( z_{\mathcal{C}}^{(i, t)} \), where \( t \) indexes the CMR frames. For ECG, each token represents a segment of length \( p_t \) from a single lead. To standardize heartbeat representations across patients, we extract tokens between consecutive R-peaks for each detected heartbeat. These variable-length sequences are interpolated to a fixed length of \( T \) tokens. Finally, we compute the mean across multiple heartbeats to obtain the ECG local embeddings for patient \( i \), denoted as \( z_{\mathcal{E}}^{(i, k)} \), where \( k \) indexes the ECG time segments.

\begin{figure}[ht!]
    \centering
    \includegraphics[width=0.99\textwidth]{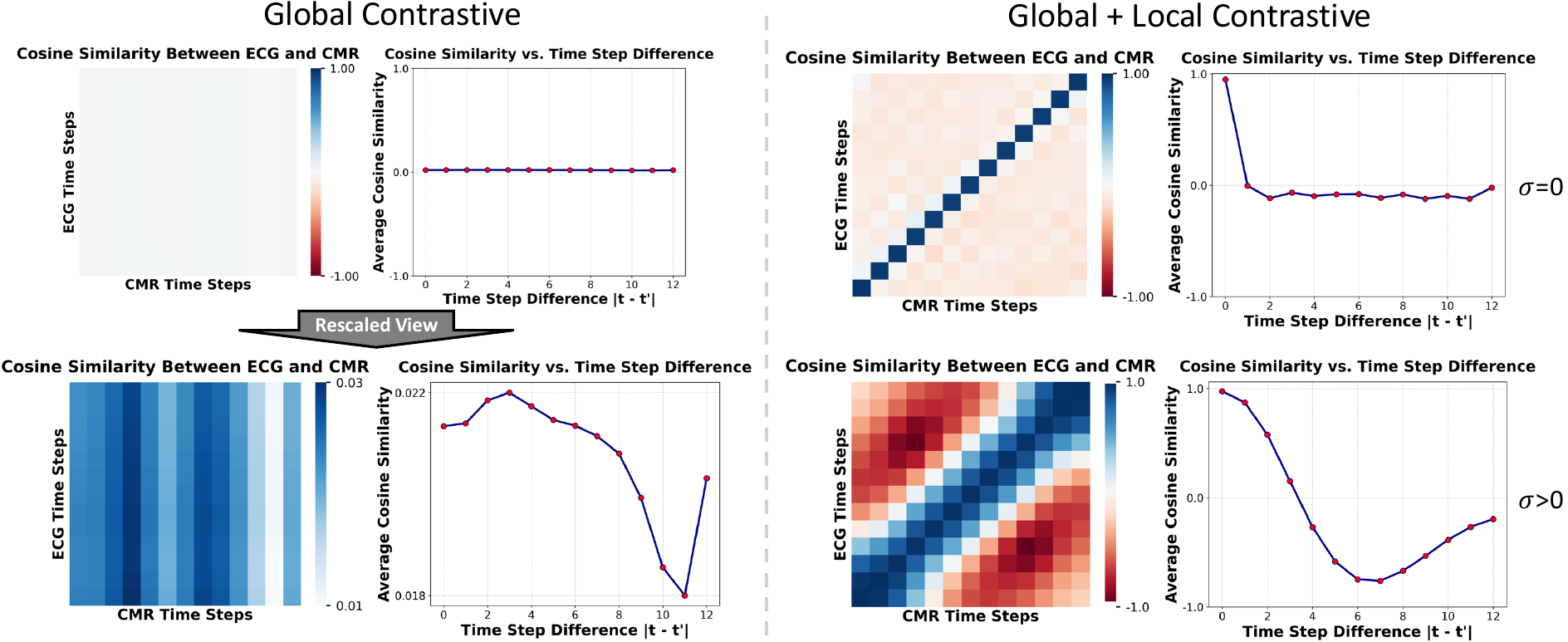}
    \caption{Impact of local contrastive loss on temporal alignment between time steps from both modalities computed during inference. The left side, using only global contrastive learning, shows no clear temporal correspondence between modalities. The right side demonstrates the effect of adding local contrastive loss (PTACL), leading to improved alignment. For $\sigma=0$, the model enforces strict one-to-one time step matching, while for $\sigma>0$, Gaussian weighting enables smoother alignment. Our method achieves temporal alignment between ECG and CMR time steps, which is absent in the global-only approach.}
    \label{fig:cosine_sim}
\end{figure}

\paragraph{Temporal Alignment and Local Loss:}
Following \cite{supcontr}, we enforce temporal alignment between \( z_{\mathcal{E}}^{(i, k)} \) (ECG) and \( z_{\mathcal{C}}^{(i, t)} \) (CMR) local embeddings, using symmetrical alignment matrix \(\mathcal{P}_{k,t}\):
\begin{equation}
    \mathcal{P}_{k,t} \propto 
    \begin{cases}
    \quad \quad \quad \delta_{k,t} & \sigma = 0 \quad (\text{hard alignment})\\
    \exp\Bigl(-\frac{1}{2}\Bigl(\frac{\text{dist}_{k,t}}{\sigma}\Bigr)^2\Bigr) &  \sigma > 0 \quad (\text{soft alignment})
    \end{cases}
\end{equation}
Here \(\sum_{t=1}^{T} \mathcal{P}_{k,t} = 1\) and \(\delta_{k,t}\) ensures exact matches for \(\sigma = 0\), while for \(\sigma > 0\), alignment is softened using a Gaussian weighting based on the normalized temporal distance \(\text{dist}_{k,t}\) (Fig.~\ref{fig:cosine_sim}).

The local contrastive loss is defined symmetrically for 
\(\text{ECG} \!\to\! \text{CMR}\) and \(\text{CMR} \!\to\! \text{ECG}\) directions. The local per-patient, per-segment loss for \(\text{ECG} \!\to\! \text{CMR}\):
\begin{equation}
\ell_{\text{ECG}\to \text{CMR}}^{(i,k)}
=
-\sum_{t=1}^{T}
\mathcal{P}_{k,t}\,
\log\!\left[
  \frac{
    \exp\Bigl(\cos(z_{\mathcal{E}}^{(i,k)},\,z_{\mathcal{C}}^{(i,t)})/\tau\Bigr)
  }{
    \sum_{t'=1}^{T}
    \exp\Bigl(\cos(z_{\mathcal{E}}^{(i,k)},\,z_{\mathcal{C}}^{(i,t')})/\tau\Bigr)
  }
\right].
\end{equation}
The total local contrastive loss for \(\text{ECG} \!\to\! \text{CMR}\) is obtained by averaging \(\ell_{\text{ECG} \!\to\! \text{CMR}}^{(i,k)}\) over all ECG segments \(k\) and patients in the batch. The final local contrastive loss \(\mathcal{L}_{\text{local}}\) is a weighted combination of both directions.

\subsubsection{Total Loss:}
The final objective combines both global and local contrastive losses to ensure alignment at both patient and temporal levels. The total loss is:
\begin{equation}
L = \mathcal{L}_{\text{global}} + \beta \cdot \mathcal{L}_{\text{local}}
\end{equation}
where \( \beta \) controls the relative contribution of the local contrastive loss.

\section{Experiments}
We evaluate our model using paired 12-lead resting ECG and short-axis CMR data from the UK Biobank \cite{ukbiobank}, with 27,951 patients for training and 5,991 for test. CMR sequences contain 50 frames, with the middle short-axis slice cropped around the cardiac center (\(84 \times 84\)) and min-max normalized. ECGs are 10-second, 12-lead recordings sampled at 500 Hz. ECGs are preprocessed via an 8-level wavelet transform, following \cite{MaeFe, ecg_filter}, for baseline wander and noise removal and followed by z-score normalization. CMR augmentations include random resized cropping, time sampling, rotation, flipping, color jitter, Gaussian blurring, and noise, while ECG augmentations include Fourier surrogate transformations, random cropping (2500 points), jittering, and rescaling. We evaluate 10 CMR-derived cardiac phenotypes \cite{Bai2020}: LV/RV end-diastolic and end-systolic volumes (EDV, ESV), stroke volume (SV), ejection fraction (EF), LV mass (M), and LV cardiac output (CO). 
%Additionally, we evaluate retrieval performance for 4 cardiovascular diseases: myocardial infarction (I21), cardiomyopathy (I42), atrial fibrillation (I48), and heart failure (I50).

\subsection{Retrieval of Similar Patients}
To assess the clinical utility of our learned ECG embeddings, we test their ability to retrieve patients with similar CMR-derived phenotypes. For each query ECG, we rank all patients in the CMR embedding database using cosine similarity. A match is counted if the candidate’s phenotype is within \(\pm 0.5\,\sigma\) of the query’s value, where \(\sigma\) is the standard deviation of that phenotype across test data. We compare two embedding strategies: a baseline model (global) and PTACL (global+local). While our global model already demonstrates strong retrieval performance, incorporating a local loss further improves precision and ranking. This enhancement is clinically significant as it enables more accurate patient stratification. Table~\ref{tab:retrieval_results} summarizes Precision$@k$ (P$@k$), mean rank (MnR), and median rank (MdR) for key heart phenotypes. For reference, values in parentheses indicate performance when retrieving $k$ random patients, reported only for $k=1$ and $k=10$ to save space.

\begin{table}[ht!]
    \centering
    \renewcommand{\arraystretch}{1.2}
    \setlength{\tabcolsep}{2pt}
    \caption{Retrieval results for CMR-derived heart phenotypes using ECG embeddings. We compare baseline (global $\triangle$) and PTACL (global+local $\square$) models, reporting performance as $\triangle$/$\square$ (random), with random retrieval in parentheses. A match is defined as a candidate whose phenotype differs from the query by no more than \(\pm0.5\,\sigma\). The inclusion of local loss in PTACL consistently improves patient retrieval with higher precision and lower rank scores across all phenotypes, demonstrating its potential to enhance clinical patient stratification.}
    \label{tab:retrieval_results}
    \resizebox{\textwidth}{!}{
    \begin{tabular}{|l|c|c|c|c|c|c|c|c|c|}
        \hline
        \rowcolor{gray!20} 
          & \textbf{P@1}$\uparrow$  & \textbf{P@3}$\uparrow$   & \textbf{P@5}$\uparrow$  & \textbf{P@10}$\uparrow$ & \textbf{P@15}$\uparrow$ & \textbf{MnR}$\downarrow$ & \textbf{MdR}$\downarrow$ \\
        \midrule
        \rowcolor{gray!10} EDV\textsubscript{LV}    & 0.492 / \textbf{0.510} (0.291) & 0.462 / \textbf{0.463} & \textbf{0.445} / 0.444 & 0.426 / \textbf{0.427} (0.287) & \textbf{0.416} / 0.415 & 3.1 / \textbf{2.9} & 2.0 / \textbf{1.0} \\
        ESV\textsubscript{LV}     & 0.501 / \textbf{0.524} (0.303) & 0.469 / \textbf{0.480} & 0.456 / \textbf{0.461} & 0.438 / \textbf{0.442} (0.284) & 0.429 / \textbf{0.432} & 3.0 / 3.0 & 1.0 / 1.0 \\
        \rowcolor{gray!10} SV\textsubscript{LV}     & 0.458 / \textbf{0.472} (0.294) & 0.417 / \textbf{0.423} & 0.404 / \textbf{0.405} & 0.387 / \textbf{0.388} (0.296) & \textbf{0.380} / 0.377 & 3.4 / \textbf{3.2} & 2.0 / 2.0 \\
        EF\textsubscript{LV}       & 0.431 / \textbf{0.436} (0.278) & 0.392 / \textbf{0.396} & \textbf{0.378} / 0.377 & 0.355 / \textbf{0.359} (0.282) & 0.349 / \textbf{0.351} & 3.9 / \textbf{3.8} & 2.0 / 2.0 \\
        \rowcolor{gray!10} CO\textsubscript{LV}       & 0.439 / \textbf{0.453} (0.295) & 0.407 / \textbf{0.417} & 0.391 / \textbf{0.394} & 0.377 / \textbf{0.378} (0.284) & 0.368 / 0.368 & 3.9 / \textbf{3.6} & 2.0 / 2.0 \\
        M\textsubscript{LV}       & 0.547 / \textbf{0.561} (0.293) & 0.516 / \textbf{0.522} & 0.498 / \textbf{0.504} & 0.479 / \textbf{0.486} (0.294) & 0.471 / \textbf{0.474} & 2.6 / 2.6 & 1.0 / 1.0 \\
        \rowcolor{gray!10} EDV\textsubscript{RV}    & 0.500 / \textbf{0.516} (0.280) & 0.467 / \textbf{0.477} & 0.450 / \textbf{0.459} & 0.433 / \textbf{0.441} (0.287) & 0.426 / \textbf{0.430} & 3.0 / \textbf{2.8} & 2.0 / \textbf{1.0} \\
        ESV\textsubscript{RV}    & 0.518 / \textbf{0.531} (0.290) & 0.485 / \textbf{0.494} & 0.471 / \textbf{0.476} & 0.455 / \textbf{0.460} (0.281) & 0.446 / \textbf{0.451} & 2.8 / \textbf{2.6} & 1.0 / 1.0 \\
        \rowcolor{gray!10} SV\textsubscript{RV}    & 0.458 / \textbf{0.473} (0.285) & 0.420 / \textbf{0.428} & 0.403 / \textbf{0.408} & 0.386 / \textbf{0.390} (0.289) & 0.380 / \textbf{0.382} & 3.8 / \textbf{3.7} & 2.0 / 2.0 \\
        EF\textsubscript{RV}    & 0.433 / \textbf{0.443} (0.288) & 0.397 / \textbf{0.405} & 0.383 / \textbf{0.390} & 0.370 / \textbf{0.373} (0.290) & 0.364 / \textbf{0.365} & 4.3 / \textbf{4.2} & 2.0 / 2.0 \\
        \hline
    \end{tabular}
    }
\end{table}

\subsection{Regression of Cardiac Phenotypes}
We evaluate the regression performance of different pre-training strategies on various cardiac phenotypes using \( R^2 \) scores. Table~\ref{tab:all_methods_comparison} presents the results for models trained on ECG, CMR, or paired ECG-CMR data. CMR single modality pre-training (SMP) reveals the upper bound of imaging information, while the lower performance of ECG SMP highlights the challenges of using ECG alone. Multimodal learning aims to reduce this gap. Among self-supervised approaches, MAE \cite{image_mae} consistently outperforms SimCLR \cite{SimCLR} on both modalities, although fully supervised single modality training remains a robust baseline. Our Global model serves as a strong baseline, already outperforming previous works. Adding a local contrastive loss (PTACL) further improves performance across all cardiac phenotypes. Importantly, PTACL achieves these gains without introducing additional learnable parameters, making it a computationally efficient enhancement.

\begin{table}[ht!]
\centering
\caption{Regression performance (\( R^2 \)) for different pre-training methods. Models are evaluated with either linear probing (LP) or full fine-tuning (FN) using attention pooling (AP). Our Global model outperforms previous multimodal approaches. Adding a local contrastive loss (PTACL) further improves results across all phenotypes without introducing additional learnable parameters.}
\label{tab:all_methods_comparison}
\setlength{\tabcolsep}{4pt}
\renewcommand{\arraystretch}{1.1}
\resizebox{\textwidth}{!}{%
\begin{tabular}{llccccccccccc}
\toprule
\rowcolor{gray!20} \textbf{Input} & \textbf{Method} & \textbf{Eval} & \textbf{LVEDV} & \textbf{LVESV} & \textbf{LVSV} & \textbf{LVEF} & \textbf{LVCO} & \textbf{LVM} & \textbf{RVEDV} & \textbf{RVESV} & \textbf{RVSV} & \textbf{RVEF} \\ 
\midrule
\multicolumn{13}{c}{\cellcolor{gray!30} \textbf{CMR \( R^2 \uparrow \)}} \\ 
CMR & Random Init  & LP  & 0.015 & 0.017 & 0.009 & 0.008 & 0.003 & 0.021 & 0.018 & 0.020 & 0.009 & 0.007 \\
\rowcolor{gray!10} CMR & SimCLR \cite{SimCLR} & LP & 0.572 & 0.612 & 0.466 & 0.513 & 0.418 & 0.628 & 0.618 & 0.681 & 0.463 & 0.520 \\ 
CMR & MAE \cite{image_mae} & LP & 0.809 & 0.772 & 0.678 & 0.489 & 0.571 & 0.824 & 0.799 & 0.784 & 0.657 & 0.492 \\ 
\rowcolor{gray!10} CMR & Supervised & FN & 0.814 & 0.789 & 0.665 & 0.470 & 0.536 & 0.823 & 0.801 & 0.808 & 0.623 & 0.473 \\ 
\midrule
\multicolumn{13}{c}{\cellcolor{gray!30} \textbf{ECG \( R^2 \uparrow \)}} \\ 
ECG  & Random Init ($\times 10^{-5}$) & LP & -4.86 & -3.89 & -7.03 & -38.8 & -7.61 & 18.3 & 0.936 & 3.61 & -9.66 & 0.983 \\
\rowcolor{gray!10} ECG & SimCLR \cite{SimCLR} & LP & 0.278 & 0.261 & 0.196 & 0.085 & 0.156 & 0.343 & 0.308 & 0.317 & 0.205 & 0.130 \\ 
ECG & MAE \cite{image_mae} & LP & 0.439 & 0.425 & 0.332 & 0.218 & 0.252 & 0.524 & 0.466 & 0.481 & 0.333 & 0.254 \\ 
\rowcolor{gray!10} ECG & Supervised & FN & 0.458 & 0.443 & 0.333 & 0.198 & 0.252 & 0.516 & 0.481 & 0.491 & 0.340 & 0.237 \\ 
\midrule
\multicolumn{13}{c}{\cellcolor{gray!30} \textbf{Multimodal (ECG+CMR) \( R^2 \uparrow \)}} \\ 
ECG & ECCL\cite{baseline1}$^*$ & FN & 0.372 & 0.348 & 0.270 & 0.176 & 0.212 & 0.449 & 0.397 & 0.397 & 0.281 & 0.203 \\

\rowcolor{gray!10} ECG & CMAE\cite{CMAE}$^\dagger$  & LP  & 0.451 & 0.380 & 0.316 & 0.103 & 0.281 & 0.536 & 0.490 & 0.445 & 0.320 & 0.129 \\

ECG & MMCL\cite{mmcl}$^\ddagger$  & FN+AP  & 0.498 & 0.497 & 0.360 & 0.245 & - & 0.597 & 0.527 & 0.534 & 0.375 & 0.248 \\

\rowcolor{gray!10} ECG & PTACL (w/o SMP)$^1$ & LP & 0.359 & 0.334 & 0.256 & 0.114 & 0.187 & 0.406 & 0.381 & 0.386 & 0.253 & 0.149 \\

ECG & PTACL (w/ CosSim++)$^2$ & LP & 0.509 & 0.496 & 0.383 & 0.252 & 0.286 & 0.614 & 0.537 & 0.553 & 0.387 & 0.295 \\

\rowcolor{gray!10} ECG & Global$^3$ & LP & 0.507 & 0.499 & 0.377 & 0.258 & 0.281 & 0.612 & 0.534 & 0.553 & 0.380 & 0.297 \\
\hline
\textbf{ECG} & \textbf{PTACL$^3$} & \textbf{LP} & \textbf{0.514} & \textbf{0.500} & \textbf{0.386} & \textbf{0.255} & \textbf{0.289} & \textbf{0.616} & \textbf{0.540} & \textbf{0.557} & \textbf{0.389} & \textbf{0.299} \\
\rowcolor{gray!10}  \textbf{ECG} & \textbf{PTACL$^3$} & \textbf{FN+AP} & \textbf{0.544} & \textbf{0.537} & \textbf{0.408} & \textbf{0.283} & \textbf{0.310} & \textbf{0.645} & \textbf{0.568} & \textbf{0.582} & \textbf{0.408} & \textbf{0.315} \\
\midrule
\multicolumn{13}{l}{$^\dagger$ Pre-trained on long-axis slices with reconstruction loss during multimodal pre-training.} \\
\multicolumn{13}{l}{$^\ddagger$ Pre-trained on three short-axis slices.} \\
\multicolumn{13}{l}{$^*$ Pre-trained on a single middle short-axis slice in time. \( R^2 \) values computed as \( r^2 \) based on reported Pearson correlation coefficients.} \\
\multicolumn{13}{l}{$^1$ Pre-trained on a single middle short-axis slice in time. w/o I phase of single modality pre-training (SMP).} \\
\multicolumn{13}{l}{$^2$ Pre-trained on a single middle short-axis slice in time. w/ local loss maximizing cosine similarity for matching time steps (no negatives).} \\
\multicolumn{13}{l}{$^3$ Pre-trained on a single middle short-axis slice in time. std not shown and in \(\sim\)0.001--0.002. w/ I phase of SMP.} 
\\
\bottomrule
\end{tabular}%
}
\end{table}

\subsection{Implementation Details}
Our approach employs transformer-based architectures for both modalities. The CMR model has 4 layers, 8 attention heads, a patch $[2, 12, 12]$, and a 90\% mask ratio (13M parameters). The ECG model has 6 layers, 8 heads, a patch 50, and a 75\% mask ratio (19M parameters). Both use a 1-layer, 4-head decoder. During multimodal pre-training, the first two layers are frozen, leaving 6.3M and 12.6M trainable parameters for CMR and ECG, respectively. The contrastive loss follows the $\mathcal{L}_{out}^{sup}$ variant from \cite{supcontr}. We use $T\!=\!13$, \( \beta\!=\!1.0 \). Regression is performed with \( \sigma\!=\!0.1 \) and global/local temperatures of 0.1/1.0, retrieval uses \( \sigma\!=\!0.0 \) with temperatures of 0.07/0.07. Linear probing is performed with ordinary least squares regression. 

%Results are averaged over five random seeds.

\section{Conclusion and Discussion}
We introduced PTACL, a contrastive learning framework that enhances ECG representations through patient-level and time step alignment with CMR. Unlike prior methods focused solely on global alignment, PTACL captures finer temporal correspondences, leading to better retrieval and phenotype regression, all without adding extra learnable parameters. Our approach relies on only a single middle short-axis slice from the CMR, demonstrating that minimal imaging data can still significantly enrich ECG.
% However, PTACL relies on paired ECG-CMR data, and our evaluation is limited to a single cohort. Future work could explore generalization to external datasets, extend PTACL to other modalities (e.g., ultrasound), or reduce reliance on paired multimodal data.
However, PTACL depends on paired ECG-CMR data, which may limit its applicability. Future work could extend PTACL to additional modalities, such as ultrasound, explore tri-modal learning, or develop strategies to mitigate the reliance on paired multimodal datasets.

\begin{credits}
\subsubsection*{\ackname}
This research has been conducted using the UK Biobank Resource under Application Number 87802. This work is funded by the European Research Council (ERC) project Deep4MI (884622). A. S. is funded via the EVUK program ("Next-generation Al for Integrated Diagnostics”) of the Free State of Bavaria.

\subsubsection*{\discintname}
The authors have no competing interests to declare that are relevant to the content of this article.
\end{credits}

%
% ---- Bibliography ----
%
% BibTeX users should specify bibliography style 'splncs04'.
% References will then be sorted and formatted in the correct style.
%

\end{document}